\documentclass[prl,twocolumn,showpacs]{revtex4}
\usepackage{graphicx}
\usepackage{times}
\usepackage{dcolumn}
\usepackage{bm}

\def\be{\begin{equation}}
\def\ee{\end{equation}}
\def\bea{\begin{eqnarray}}
\def\eea{\end{eqnarray}}
\def\bse{\begin{subequations}}
\def\ese{\end{subequations}}

\def\be{\begin{eqnarray}}
\def\ee{\end{eqnarray}}

\newcommand{\e}{\epsilon}

\begin{document}

\title{Testable signatures of quantum non-locality in a two-dimensional
chiral $p$-wave superconductor }
\author{Sumanta Tewari$^{1}$}
\author{Chuanwei Zhang$^{1}$}
\author{S. Das Sarma$^{1}$}
\author{Chetan Nayak$^{2,3}$}
\author{Dung-Hai Lee$^{4,5}$}
\affiliation{$^{1}$Condensed Matter Theory Center, Department of Physics, University of
Maryland, College Park, MD 20742\\
$^{2}$Microsoft Station Q, CNSI Building, University of California, Santa
Barbara, CA 93108\\
$^{3}$Department of Physics and Astronomy, University of California, Los
Angeles, CA 90095-1547\\
$^{4}$Department of Physics, University of California at Berkeley, Berkeley,
CA 94720 \\
$^{5}$ Material Science Division, Lawrence Berkeley National Laboratory,
Berkeley, CA 94720, USA }

\begin{abstract}
A class of topological excitations -- the odd-winding number vortices -- in
a spinless 2D chiral $p$-wave $(p_x+ip_y)$ superconductor trap Majorana
fermion states in the vortex cores. For a dilute gas of such vortices,
the lowest energy \textit{fermionic} eigenstates
are intrinsically non-local. We predict two testable signatures of this
unusual quantum non-locality in quasiparticle tunneling experiments. We
discuss why the associated teleportation-like phenomenon does not imply the
violation of causality.
\end{abstract}

\maketitle




\emph{Introduction:} The intriguing physics of the Majorana fermion zero
modes at the vortex cores of a spinless $p_x+ip_y$-wave
superconductor/superfluid has recently attracted considerable attention \cite%
{Read, Ivanov, Stern, Stone, Sumanta1, Sumanta2, Sumanta3,
Gurarie1}. This is because these zero modes endow the vortices with
non-Abelian braiding statistics, and the braiding of such vortices
\cite{Nayak} can, in principle, be exploited to build a
fault-tolerant topological quantum computer (TQC) \cite{Kitaev}.
Recently there has been considerable evidence
\cite{Kapitulnik,Harlingen} that the symmetry of the superconducting
order parameter in strontium ruthenate (Sr$_2$RuO$_4$)
\cite{Maeno,Mackenzie} is spin-triplet $p_x+ip_y$. It has been
proposed in Ref. \onlinecite{Sumanta1} that when a thin film of this
material is subjected to a magnetic field exceeding $\sim 200$ G,
the most
energetically favorable vortices enclose flux $hc/4e$ rather than $hc/2e$,
and at the core of such a half vortex there is a Majorana zero mode. For
other proposals for the superconducting state in this material,
see Refs.~\onlinecite{Zutic, Machida}. Some heavy fermion materials,
such as UPt$_3$, are likely to be spin-triplet superconductors as well.
Moreover, with the recent observations of $p$-wave Feshbach resonances in
spin-polarized $^{40}$K and $^{6}$Li atoms in optical traps \cite%
{Regal,Ticknor}, these systems are promising candidates for realizing $p$%
-wave condensates as well \cite%
{Gurarie, Cheng}.
Thus,
vortices satisfying non-Abelian statistics are tantalizingly close to
experimental reach.
In the following, for brevity, we will use the
term ``superconductor'' to indicate both superconductor and
superfluid.

An inevitable consequence of the existence of Majorana zero modes is that
the fermion quasiparticle excitations are inherently non-local.
The zero energy bound
state at the vortex core
is described by the Majorana fermion operator
$\gamma^{\dagger}=\gamma$ \cite{Read}. Other bound states
are separated by an energy gap $\omega_0 \sim
\frac{\Delta_0^2}{\epsilon_{\mathrm{F}}}\ll \Delta_0$, where
$\Delta_0$ gives the gap in the bulk and $\epsilon_{\mathrm{F}}$ is
the Fermi energy.
 If the system has two Majorana zero modes, $\gamma_1$ and
 $\gamma_2$,
one can define a composite operator $%
q^{\dagger}= ({\gamma_1 +i\gamma_2})/\sqrt{2}$, and its hermitian conjugate $%
q$, which satisfy Fermi anticommutation relations. If an electron is
injected into the system with an energy $\ll\omega_0$, it can only
go to the excited state described by $q$ and $q^{\dagger}$.
This intrinsic non-locality of the quasiparticle wavefunction
was first discussed in Refs.~\onlinecite{Semenoff-unpub,Semenoff} for a 1D
quantum wire embedded in a $p$-wave superconductor. 
In this paper, we use it to make interesting predictions for
quasiparticle tunneling experiments in a 2D $p_x+ip_y$
superconductor.
To distinguish the non-locality discussed above from that in a quantum
double well problem, let $|l\rangle$ and $%
|r\rangle$ represent the two states localized in the left and the right
well. In the presence of quantum tunneling, $-t(c^\dagger_lc_r+\mathrm{{h.c.}%
)}$, the ground and the excited states are given by the symmetric and the
anti-symmetric states, $|s\rangle=\frac{1}{\sqrt{2}}(c^\dagger_l+c^%
\dagger_r)|0\rangle$ and $|a\rangle=\frac{1}{\sqrt{2}}(c^\dagger_l-c^%
\dagger_r)|0\rangle$, respectively. Thus, it might appear that in this case
also the quasiparticle operators are delocalized between the two wells even
when they are very far separated.
However, if we inject an electron into the left well,
it will enter into the linear combination
of $|s\rangle$ and $|a\rangle$, the state localized in the left well \cite%
{Semenoff-unpub,Semenoff}. Only after a time of the order of $\hbar$ divided
by the tunnel splitting
does the electron acquire appreciable probability of appearing at
the right well.
Further, because
of the
exponentially small tunnel splitting, by adding a small perturbation, e.g., $%
\epsilon_Lc^\dagger_Lc_L$, to the Hamiltonian, the eigen
wavefunction will become localized near the left or the right well.
Motivated by these considerations we consider the
following thought experiments to test the explicit non-locality of a
quasiparticle excitation in the Majorana fermion system.\newline

\emph{Teleportation:}
\begin{figure}[tbp]
\includegraphics[angle=0,scale=0.5]{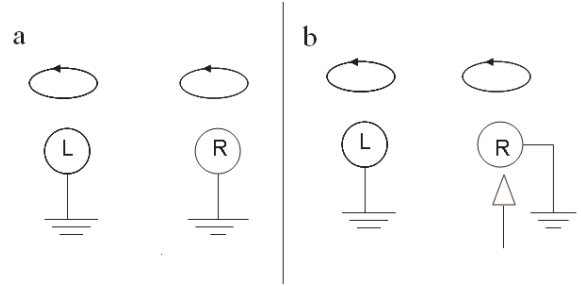}
\caption{(a) Two quantum dots in tunneling contact with two vortices. (b) A
quantum dot and a STM tip in tunneling contact with two vortices.}
\end{figure}
We consider the set up of Fig.~1(a) where two voltage biased quantum dots
are in tunneling contact with two non-Abelian vortices of a $p_x+ip_y$
superconductor. The effective Hamiltonian describing the system at energy $<<%
{\omega}_0$ is given by,
\begin{eqnarray}
H&&=\sum_{\alpha=L,R}\epsilon_\alpha\psi^\dagger_\alpha\psi_\alpha+iJ%
\gamma_L\gamma_R -\sqrt{2}t_L\gamma_L(\psi^{\dagger}_L-\psi_L)  \nonumber \\
&&-\sqrt{2}t_R\gamma_R(\psi^{\dagger}_R-\psi_R)  \label{h1}
\end{eqnarray}
Here, $\gamma_L$ and $\gamma_R$ are the two Majorana fermion operators
and $\psi_L$ and $\psi_R$ are the
electron annihilation operators for the lowest energy state of the quantum
dots. The parameter $J$ describes the hybridization between the Majorana
fermion modes. It is exponentially small if the distance between the
vortices is large.

After introducing fermion operators $q^{\dagger}=({\gamma_R+i\gamma_L})/%
\sqrt{2}$, $q=({\gamma_R-i\gamma_L})/\sqrt{2}$, and performing the local
gauge transformation $\psi_L\rightarrow i\psi_L$, Eq.~(\ref{h1}) becomes
\cite{Demler}
\begin{eqnarray}
H&&=\sum_{\alpha=L,R}\epsilon_\alpha\psi^\dagger_\alpha\psi_\alpha+Jq^%
\dagger q+\sum_{\alpha=L,R}t_{\alpha}(q^{\dagger}\psi_{\alpha}+\mathrm{h.c.})
\nonumber \\
&&-t_{L}(\psi^{\dagger}_{L}q^{\dagger}+\mathrm{h.c.})+t_{R}(\psi^{%
\dagger}_{R}q^{\dagger}+\mathrm{h.c.}).  \label{h2}
\end{eqnarray}
In this form, it is clear that the tunneling terms only conserve charge
modulo $2e$. (The extra or deficit charge is taken away or provided by the
superconducting condensate.) The lowest energy Hilbert space upon which Eq.~(%
\ref{h2}) acts is eight dimensional. Its basis states are labeled by $%
|n_L,n_q,n_R\rangle$, where $n_{L,q,R}=0,1$ denote the number of particles
in the single particle states associated with the $\psi_L$, $q$ and $\psi_R$
operators, respectively. Because of charge conservation modulo $2e$, this
eight dimensional Hilbert space decouples into two four dimensional ones
spanned by $\{|1,0,0\rangle,|0,1,0\rangle,|0,0,1\rangle,|1,1,1\rangle\}$ and
$\{|0,1,1\rangle,|1,0,1\rangle,|1,1,0\rangle,|0,0,0\rangle\}$, respectively.
In the demonstration of the teleportation phenomenon, we shall consider one
extra electron at the left dot at time zero and see how it is transported to
the right dot at a later time. For that purpose, we only need to consider
the subspace where the number of electrons is odd. In this subspace the
Hamiltonian is a $4\times 4$ matrix given by
\begin{eqnarray}
H=\pmatrix{\e_L&t_L&0&-t_R\cr t_L&J&t_R&0\cr 0&t_R&\e_R&-t_L\cr
-t_R&0&-t_L&J+\e_L+\e_R}.  \label{hodd}
\end{eqnarray}

To simplify the calculations, in the following we shall consider the
situation where $J\rightarrow 0^+$, $t_L=t_R=t$, $\epsilon_R=0$, and study
the eigenstates of Eq.~(\ref{h1}) as a function of $\epsilon_L$. Physically,
we expect that when $\epsilon_L$ is large and negative, the extra electron
initially at the left dot will stay there. On the contrary, by tuning $%
\epsilon_L$ to zero, the extra electron can delocalize by tunneling to the
right dot via the conduit provided by the extended $q$-state. Notice that
this is very different from ordinary quantum tunneling of the electron which
is non-zero only when $J$ is appreciable.

A simple calculation shows that if the initial state is $|1,0,0\rangle$, the
probability of observing an electron at the left and the right dots at a
later time $T$ is given by
\begin{eqnarray}
P_L(T)&=&|\langle 1,0,0|e^{-iHT}|1,0,0\rangle|^2+|\langle
1,1,1|e^{-iHT}|1,0,0\rangle|^2  \nonumber \\
&=&1-{\frac{2t^2}{4t^2+\epsilon_L^2}}\left[1-\cos{(T\sqrt{4t^2+\epsilon_L^2})%
}\right]  \nonumber \\
P_R(T)&=&|\langle 0,0,1|e^{-iHT}|1,0,0\rangle|^2+|\langle
1,1,1|e^{-iHT}|1,0,0\rangle|^2  \nonumber \\
&&=\sin^2(Tt).  \label{hy}
\end{eqnarray}
Eq.~(\ref{hy}) predicts that $P_L\rightarrow 1$ for large negative $%
\epsilon_L$. Therefore, as expected, the electron initially on the left dot
will remain there. What is more important is the fact that the probability
of finding an electron on the right quantum dot is \textit{independent of $%
\epsilon_L$}!

If one separates $P_R(T)$ into the sum of the following probabilities,
\begin{eqnarray}
&&P_{R1}(T)=|\langle 0,0,1|e^{-iHT}|1,0,0\rangle|^2  \nonumber \\
&&P_{R2}(T)=|\langle 1,1,1|e^{-iHT}|1,0,0\rangle|^2,
\end{eqnarray}
one would obtain,
\begin{eqnarray}
P_{R1}(T)&=& {\frac{4t^2}{4t^2+\epsilon_L^2}}\sin^2(Tt)\sin^2\Big({\frac{T%
\sqrt{4t^2+\epsilon_L^2}}{2}}\Big)  \nonumber \\
P_{R2}(T)&=&\sin^2(Tt)-P_{R1}(T).  \label{hy1}
\end{eqnarray}
The first term, $P_{R1}(T)$, is the probability of transporting the electron
from the left to the right dot \textit{without disturbing the superconductor}%
. For $\epsilon_L=0$ it is equal to
$P_{R1}(T)= \sin^4(Tt)$,
hence it equals to unity when $T={\pi/ 2t}.$ The fact that the above time is
finite even in the limit of $J\rightarrow 0^+$ suggests that it is
independent of the (large) separation between the vortices, thus justifying
the word ``teleportation''.

Note that the \textit{total} probability of observing an electron at the
right dot, Eq.~(\ref{hy}), is independent of what's being done at the left
dot. Since there is no way to distinguish between the processes responsible
for $P_{R1}$ and $P_{R2}$ by a measurement \textit{only on the right dot},
there is no way to know if the electron has been teleported from the left
dot or has arisen out of the
condensate. As a result, no information can be sent from the left to
the right dot, hence causality is maintained. To prove the existence
of teleportation, it is necessary to differentiate the processes
responsible for $P_{R1}$ and $P_{R2}$. For example, one could
perform a coincident measurement of both the occupation numbers of
the left and the right dots at time $t$. This measurement will give
zero for the
process responsible for $P_{R1}$ and non-zero for that responsible for $%
P_{R2}$. However, since a `classical' exchange of information (the result of
the coincident measurement) is necessary, there is no superluminal transfer
of information in the observation of the teleportation effect. This
experiment is possible in principle, but will be difficult in practice. In
the following, we describe a much easier STM experiment which reveals the
non-local nature of the $q$-quasiparticle state discussed earlier. \newline

\emph{Action over distance:} Now let us consider the set up in Fig.~1(b)
where a voltage-biased quantum dot is in tunneling contact with the left
vortex, while an STM tip probes the differential conductance dI/dV of the
right dot. The purpose is to study the effect of biasing the left dot on the
tunneling curve at the right dot. Again one might expect a non-zero effect
due to the extended q-state.

In computing the STM spectral function, we use the formulae,
\begin{eqnarray}
A({\omega})=\sum_{\alpha}|\langle\alpha_{-\eta}|\psi^\dagger_R|0_\eta%
\rangle|^2 \delta({\omega}-E^{-\eta}_\alpha+E^{\eta}_0)
\end{eqnarray}
for ${\omega}>0$, and
\begin{eqnarray}
A({\omega})=\sum_{\alpha}|\langle\alpha_{-\eta}|\psi_R|0_\eta\rangle|^2
\delta({\omega}-E^{-\eta}_\alpha+E^{\eta}_0)
\end{eqnarray}
for ${\omega}<0.$ Here $\eta=\pm 1$ labels the even and odd particle number
subspaces respectively, 
and $|\alpha_\eta\rangle, E^\eta_\alpha$ are the eigenstates and the
eigenenergies of Eq.~(\ref{h2}) ($\alpha=0$ indicates the ground state). In
the explicit calculation, we diagonalize the following $8\times 8$
Hamiltonian matrix,
\begin{eqnarray}
H=\pmatrix{H_1&0\cr 0&H_2},
\end{eqnarray}
where $H_1$ is the $4\times 4$ matrix given in Eq.~(\ref{hodd}) and $H_2$ is
given by,
\begin{eqnarray}
H_2=\pmatrix{\e_R+J&t_L&0&-t_R\cr t_L&\e_L+\e_R&t_R&0\cr
0&t_R&\e_L+J&-t_L\cr -t_R&0&-t_L&0}.
\end{eqnarray}
In this $8-$dimensional space $\psi^\dagger_R$ is represented by
\begin{eqnarray}
\psi^\dagger_R=\pmatrix{0&C_{oe}\cr C_{eo}&0 }
\end{eqnarray}
where
\begin{eqnarray}
C_{oe}=\pmatrix{0&0&0&0\cr0&0&0&0\cr0&0&0&1\cr0&0&1&0}, C_{eo}=%
\pmatrix{0&-1&0&0\cr -1&0&0&0\cr0&0&0&0\cr 0&0&0&0}
\end{eqnarray}

For $J=0$, straightforward computation gives
\begin{eqnarray}
A({\omega})={\frac{1}{2 \pi}} {\frac{\Gamma [2t^2+ \Gamma^2 +|{\omega}|(|{%
\omega}|-2t)]}{({\omega}^2+\Gamma^2)[(|{\omega}|-2t)^2+\Gamma^2]}}, ,
\label{aw}
\end{eqnarray}
where $\Gamma$ is a Lorentzian broadening of the energy levels. The fact
that the result is independent of $\epsilon_L$ is expected because, for zero
$J$, there is no communication between the two Majorana fermion states
associated with $\gamma_L$ and $\gamma_R$. As a result, Eq.~(\ref{h1})
decouples into two independent quantum dots in tunneling contact with two
independent vortices. Because of this, the tunneling spectrum of the right
dot is independent of
the left dot.

Next let us consider the case of non-zero $J$. In this case, the expression
for $A({\omega})$ is lengthy and we shall just present the numerical results
for $J=0.5 t$. In Fig.~2 the dashed curve is the tunneling spectrum for $%
\epsilon_L=0$ while the solid curve is that for $\epsilon_L=t$. In
constructing this figure we used $\Gamma=0.15 t$. It is clear that the
tunneling spectrum at the right dot can be \textit{\ qualitatively altered}
by tuning $\epsilon_L$ only on the left dot. The most surprising result is
that, when $\epsilon_L\ne 0$, there is a discontinuity in $A({\omega})$ at ${%
\omega}=0$!
\begin{figure}[tbp]
\includegraphics[scale=0.5]{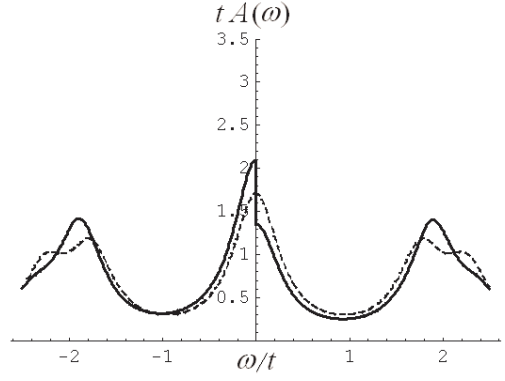}
\caption{The tunneling curves for $J=0.5t$. Dashed curve: $\protect\epsilon%
_L=0$, solid curve: $\protect\epsilon_L=t$.}
\label{asymm}
\end{figure}
The presence of zero energy spectral weight is due to the relation:
\begin{eqnarray}
E_0^{-\eta}=E_0^\eta.  \label{se}
\end{eqnarray}
The discontinuity arises from the fact that
\begin{eqnarray}
|\langle 0_{-\eta}|\psi^\dagger_R|0_\eta\rangle|^2\ne |\langle 0
_{-\eta}|\psi_R|0_\eta\rangle|^2.  \label{neq}
\end{eqnarray}
Eq.~(\ref{se}) is due to the fact that, for $\epsilon_R=0$, $H_1$ and $H_2$
have identical eigenvalues. As for Eq.~(\ref{neq}), it can be shown
analytically that
\begin{eqnarray}
&&|\langle 0_{-\eta}|\psi^\dagger_R|0_\eta\rangle|^2- |\langle 0
_{-\eta}|\psi_R|0_\eta\rangle|^2  \nonumber \\
=&&|\langle 0_{\eta}|\psi_R|0_{-\eta}\rangle|^2- |\langle 0
_{-\eta}|\psi_R|0_\eta\rangle|^2  \nonumber \\
=&&{\frac{J\epsilon_L}{\sqrt{16 t^4+\epsilon_L^2(4t^2+J^2)}}}
\end{eqnarray}
which yields zero (no discontinuity) if either $J$ or $\epsilon_L$ is zero.

A discontinuity in the tunneling spectrum is very unusual. The presence of a
non-zero tunneling spectral weight at ${\omega}=0$ requires the absence of
an energy gap. This occurs, for example, in metals and Anderson insulators.
The discontinuity in the spectral weight at ${\omega}=0$ then requires $%
|\langle 0_{N-1}|\psi_R|0_N\rangle|\ne |\langle
0_N|\psi_R|0_{N+1}\rangle|$, where $|0_M\rangle$ is the M-particle
ground state. This would not occur in a metal or an Anderson
insulator.

\emph{Experiments in atomic superfluid:}
The schemes proposed above for testing the non-local properties of
the Majorana zero modes can be implemented in a
$(p_{x}+ip_{y})$-wave atomic
superfluid through a recently proposed laser probing scheme \cite%
{Torma,Bruun}. This scheme is similar to the tunneling microscopy of a
superconductor in that it relies on induced tunneling between a superfluid
and a normal phase. Here, we assume that the atoms in the superfluid are
prepared in a hyperfine state $\left\vert \downarrow \right\rangle $, while
the atoms in the other hyperfine state $\left\vert \uparrow \right\rangle $
do not participate in the $p$-wave pairing, and therefore constitute the
normal phase. The interference between the superfluid and normal phases can
be realized through a two-photon Raman process that couples the two states $%
\left\vert \downarrow \right\rangle $ and $\left\vert \uparrow
\right\rangle $ of atoms with two local laser fields. An optical
dipole trap with spin-dependent potential depth can be used to hold
the normal atoms near the vortex cores. The bias voltage
$\varepsilon _{L}$ can be adjusted by varying the intensity of the
optical dipole trap, which changes the energy of the normal atoms
with respect to atoms in the superfluid. The number of atoms in the
normal phase can be measured through fluorescence signals of
resonance lasers. The tunneling strengths $t_{L}$ and $t_{R}$ can be
adjusted by varying the intensities of the Raman lasers fields,
which changes the coupling strength between the two hyperfine
states. In the experiment, to detect the action over distance, the
spectral function $A\left( \omega \right) $ can be measured by
varying the detuning of the laser fields to the atomic transition
$\delta =\omega _{L}-\omega _{A}$ \cite{Torma,Bruun}, where $\omega
_{L}$ is the frequency difference between the two laser fields for
the two-photon Raman process, and $\omega _{A}$ is the frequency
splitting between the two hyperfine levels $\left\vert \downarrow
\right\rangle $ and $\left\vert \uparrow \right\rangle $. In this
way, the proposed experiments to test the quantum teleportation and
the action over distance can be performed on an atomic superfluid as
well. The mathematical illustrations, including the form of the
Hamiltonian and the measurable quantities, remain the same as in an
electronic superconductor.

\emph{Experimental feasibility:} Here we comment on the cut-off
energy scales below which the interesting effects discussed above
would be observable. The tunneling amplitudes $t_L$ and $t_R$ (the
bias voltages) need to be much
smaller than $\omega_0$. For Sr$_2$RuO$_4$ (assuming it is a spin-triplet $p$%
-wave superconductor) using $T_c\sim 1.5$ K and the effective fermi
temperature $T_F\sim 50$ K \cite{Maeno}, this requires $t_R, t_L \ll
\omega_0\sim 50$ mK. In the case of Feshbach superfluids, where $%
\Delta_0\sim \epsilon_F$ in the unitary regime \cite{Greiner}, $%
\omega_0\sim\epsilon_F$ and the effects due to the Majorana fermions
should be observable below the scale of the Fermi temperature
itself. It is also important that the experiment be done on a time
scale short compared to the decoherence time. The relevant processes
are (1) thermal excitation of a fermion zero mode to a higher-energy
state, and (2) quantum tunneling of the zero mode from the vortex
probed in our experiment to some other vortex induced by disorder.
Process (1) will occur with probability $\sim e^{-{\omega_0}/T}$.
For $T\ll 50$mK, this will be small in Sr$_2$RuO$_4$ (note that
disorder may also suppress the excitation gap $\omega_0$ locally,
which would require the temperature to be reduced further); in a
cold atom system, we merely need $T\ll\epsilon_F$. The second
process can be avoided by performing the experiment with vortices
which are well-isolated from any other vortices (which can be
detected by various imaging techniques).

To conclude, the observation of our predicted nonlocal quantum entanglement
behavior
would be direct evidence supporting the existence of
non-Abelian topological anyons in 2D chiral p-wave superfluids and
superconductors.

We thank V. W. Scarola for discussions. This research has been
supported by ARO-DTO under grant W911NF-04-1-0236, ARO-LPS,
Microsoft Corporation, and the NSF under grants DMR-0411800 and DMR
0456669. ST thanks the Institute for Complex Adaptive Matter (ICAM)
for an exchange fellowship and KITP, UCSB for hospitality. DHL was
supported by the Director, Office of Science, Office of Basic Energy
Sciences, Materials Sciences and Engineering Division of the U.S.
Department of Energy under Contract No. DE-AC02-05CH11231.

\vskip -6mm 

\end{document}